\documentclass[12pt]{article}
\pdfoutput=1

\usepackage{color}
\usepackage{amsmath}
\usepackage{amsfonts}
\usepackage{amssymb}
\usepackage{caption}
\usepackage{graphicx}
\usepackage{slashed}            
\usepackage{subfig}             
\usepackage{xspace}				
\usepackage{cite}
\usepackage[english]{babel}
\usepackage{fancyhdr}
\usepackage{amsmath}
\usepackage{amssymb}
\usepackage{amsfonts}
\usepackage{psfrag}
\usepackage[applemac]{inputenc}
\usepackage{graphicx}
\usepackage{braket}
\usepackage{breqn}
\usepackage{feynmp}
\usepackage{feynmp-auto}
\usepackage{subfig}
\usepackage[export]{adjustbox}
\usepackage[toc,page]{appendix}
\usepackage{mathtools} 
\usepackage{float}
\usepackage[mathscr]{euscript}
\usepackage{breqn}

\newcommand{\arXivold}[2]{\href{http://arxiv.org/pdf/#1}{{\tt #2/#1}}}

\usepackage[
	colorlinks=true,
	citecolor=black,
	linkcolor=black,
	urlcolor=blue,
	hypertexnames=false]{hyperref}

\numberwithin{equation}{section} 

\begin{document}
\begin{titlepage}

\begin{center}

	{	
		\LARGE \bf 
		The Scale of New Physics from the \\ Higgs Couplings to $gg$
	}
	
\end{center}
	\vskip .3cm

\begin{center} 
{\bf \  Fayez Abu-Ajamieh\footnote{\tt
		 \href{fayezajamieh@IISc.ac.in}{fayezajamieh@IISc.ac.in}
\setcounter{footnote}{0}
		 }
		} 
\end{center}

\begin{center} 

	{Center for High Energy Physics (CHEP), Indian Institute of Science (IISc) \newline
	C.V. Raman Avenue, Bangalore 560012 - India}

\end{center}


\centerline{\large\bf Abstract}

\begin{quote}
Measuring the Higgs couplings accurately at colliders is one of the best routes for finding physics Beyond the Standard Model (BSM). If the measured couplings deviate from the SM predictions, then this would give rise to energy-growing processes that violate tree-level unitarity at some energy scale, indicating new physics. In this paper, we extend previous work on unitarity bounds from the Higgs potential and the Higgs couplings to vector bosons, the top quark, $\gamma\gamma$ and $\gamma Z$; to the Higgs coupling to $gg$. We show that the scale of new physics could be as low as $\sim 10$ TeV, which although beyond the reach of the LHC, is well within the reach of the future $100$-TeV collider, and can also be probed indirectly in the High Luminosity LHC (HL-LHC).

\end{quote}

\end{titlepage}


\section{Introduction}
Unitarity arguments have long been used to investigate the possibility of new physics. For any Quantum Field Theory (QFT) to be consistent, it has to preserve unitarity to all scales, otherwise, the theory is incomplete, which is usually a sign of new physics at or below the scale at which unitarity is violated.

A classical example of the use of unitarity to predict new physics is the prediction of the Higgs boson from the unitarity violation of the scattering amplitude of the longitudinal modes of gauge bosons \cite{Lee:1977yc, Lee:1977eg}. There, Lee, Quigg, and Thacker showed that the scattering amplitude of the longitudinal modes of gauge bosons is quadratically divergent with the center of momentum energy and that the amplitude can be unitarized by including the Higgs exchange diagram which exactly cancels the divergence, rendering a finite amplitude at all energies. The requirement that unitarity be respected translates into an upper bound on the Higgs mass
\begin{equation}\label{eq:HiggsUpperBound}
M_{H} \leq \Big(\frac{8\sqrt{2}\pi}{3G_{F}} \Big)^{\frac{1}{2}} \simeq 1 \hspace{1 mm} \text{TeV},
\end{equation}
and indeed, the Higgs was discovered with a mass of $125$ GeV \cite{ATLAS:2012yve, CMS:2012qbp}.

In the same spirit, unitarity arguments can be utilized to set an upper limit on the scale of new physics. The reason is that the SM is the unique UV-complete theory with the observed particle content that can be extrapolated to arbitrarily high energies, which means that any deviation from the SM predictions will ruin this UV-completeness. This UV-incompleteness will manifest itself as energy-growing amplitudes that eventually violate unitarity at some high energy scale, signaling the onset of new physics, in exactly the same manner as Lee, Quigg, and Thacker's argument points to the upper limit on the Higgs mass, and the current level of uncertainties on the measurement of the Higgs couplings to other SM particles leaves ample room for new physics Beyond the Standard Model (BSM), which provides further motivation for this argument.

This argument was used in \cite{Chang:2019vez, Abu-Ajamieh:2020yqi} to probe the scale of new physics in the Higgs potential, as well as the Higgs couplings to the top quark and the $W/Z$, whereas the $h\gamma\gamma$ and $h\gamma Z$ sectors were investigated in \cite{Abu-Ajamieh:2021egq}. In this study, we seek to extend this argument to the $hgg$ sector. As the case with the $h\gamma\gamma$ and $h\gamma Z$ sectors, $hgg$ is loop-induced, thus we utilize the same effective coupling approach utilized in \cite{Abu-Ajamieh:2021egq}, where the loops are integrated out in order to write these couplings as tree-level couplings. Currently, experimental measurements constrain the coupling $hgg$ to be within $O(0.1)$, 

Previous studies in the literature utilized the SM Effective Field Theory (SMEFT) to investigate the couplings $hgg$ \cite{Contino:2012xk, Goertz:2014qta, Azatov:2015oxa, Deutschmann:2017qum, Deutschmann:2017biq, Grazzini:2017szg, Henning:2018kys, Ellis:2018gqa, Agrawal:2019bpm, Ellis:2020unq, Battaglia:2021nys}. However, in this paper, we adopt the completely model-independent bottom-up approach that was used in \cite{Chang:2019vez, Abu-Ajamieh:2020yqi, Abu-Ajamieh:2021egq}, which does not rely on power counting but parametrizes the scale of new physics as deviations from the SM predictions. In this approach, no assumptions are made regarding the deviations and the infinitely many unconstrained higher-order couplings, other than that they are compatible with experimental measurements. This approach avoids the SMEFT assumption of only keeping the LO operator and neglecting higher-order ones. In addition, this approach makes it easier to generalize $2 \rightarrow 2$ scattering to $n \rightarrow m$ scattering, which turns out to provide stronger bounds. 

As we show in this study, due to the loop suppression of the Higgs coupling to $ggh$, the scale of new physics is generally much higher than what was obtained from the tree-level couplings in \cite{Chang:2019vez, Abu-Ajamieh:2020yqi}, but lower than that from the $\gamma\gamma$ and $\gamma Z$ sectors \cite{Abu-Ajamieh:2021egq} due to the strong coupling. For instance, it was found in \cite{Chang:2019vez, Abu-Ajamieh:2020yqi}, that the current level of constraints on the couplings $hVV$ and $h\bar{t}t$ allows for a scale of new physics that could be as low as $\sim 3$ TeV, which is well within the reach of the HL-LHC. On the other hand, we will show in this paper that the model-independent scale of new physics from the Higgs coupling to $gg$ is generally $ \gtrsim 10$ TeV unless the deviations from the SM predictions are unnaturally large. This is beyond the reach of the HL-LHC, however, we show that one can probe these couplings indirectly, making these sectors worth investigating.

This paper is organized as follows: In Section \ref{Sec:Review}, we present a quick review of the model-independent approach and the main results found in \cite{Chang:2019vez, Abu-Ajamieh:2020yqi}. In Section \ref{Sec:hgg} we investigate new physics in the $hgg$ sector and discuss the relationship with SMEFT. We relegate the lists of full results to the appendices. We present some simple UV completions in Section \ref{sec:UVcompletion}, and finally, we present our conclusions in Section \ref{sec:conclusions}.

\section{Review of the Model-independent Approach}\label{Sec:Review}
This section is intended as a comprehensive review of the model-independent approach utilized in \cite{Chang:2019vez, Abu-Ajamieh:2020yqi, Abu-Ajamieh:2021egq}, so that the reader can obtain all salient points herein. Readers who are interested in further details regarding the scale of new physics from the Higgs potential, as well as from the Higgs interaction with the top quark, $W/Z$, $\gamma\gamma$ and $\gamma Z$ are instructed to refer to \cite{Chang:2019vez, Abu-Ajamieh:2020yqi, Abu-Ajamieh:2021egq}. 

In the literature, physics BSM is usually captured through the SMEFT approach, where one would systematically enumerate all higher-order operators that are consistent with the symmetries of the SM, suppressed by the appropriate power of the UV scale $\Lambda$. While SMEFT is quite systematic and elegant, it is nonetheless not quite model-independent, as a single UV scale of new physics is assumed in the operator expansion, which needn't be the case. In addition, only the leading order SMEFT operators are commonly retained, which implicitly assumes that the scale of new physics is high enough that higher-order operators can be safely neglected, an assumption that might not hold for all we know. Furthermore, SMEFT is not very transparent phenomenologically, as the couplings are what is measured in colliders and not the UV scale $\Lambda$. 

Given the shortcomings of SMEFT, we are thus motivated to adopt a completely model-independent approach, where the new physics in the Effective Theory (EFT) is captured through deviations in the Higgs couplings from the SM predictions. The only constraint on these deviations is that they are compatible with measurements. With this approach, we can write the effective Lagrangian of the Higgs potential and its interactions with the top quark and gauge bosons (including the effective couplings to $\gamma\gamma$ and $\gamma Z$) in the unitary gauge as follows
\begin{flalign}\label{eq:effLag1}
\mathcal{L} & = \mathcal{L}_{\text{SM}} - \delta_{3} \frac{m_{h}^{2}}{2v}h^{3} - \delta_{4} \frac{m_{h}^{2}}{8v^{2}}h^{4} - \sum_{n=5}^{\infty}\frac{c_{n}}{n!}\frac{m_{h}^{2}}{v^{n-2}}h^{n} +\nonumber \cdots \\
& + \delta_{Z1}\frac{m_{Z}^{2}}{v}h Z_{\mu}Z^{\mu} +\delta_{W1}\frac{2m_{W}^{2}}{v}hW_{\mu}^{+}W^{\mu-} + \delta_{Z2}\frac{m_{Z}^{2}}{2v^{2}}h^{2} Z_{\mu}Z^{\mu} +\delta_{W2}\frac{m_{W}^{2}}{v^{2}}h^{2}W_{\mu}^{+}W^{\mu-}   \nonumber \\
& + \sum_{n=3}^{\infty}\Big[ \frac{c_{Zn}}{n!}\frac{m_{Z}^{2}}{v^{n}}h^{n}Z_{\mu}Z^{\mu} + \frac{c_{Wn}}{n!}\frac{2m_{W}^{2}}{v^{n}}h^{n}W^{+}_{\mu}W^{-\mu} \Big] + \cdots \\
& + c^{\text{SM}}_{\gamma 1} \delta_{\gamma 1} \frac{\alpha}{\pi v} hA_{\mu\nu}A^{\mu\nu} + A_{\mu\nu}A^{\mu\nu} \sum_{n=2}^{\infty}\frac{c^{\text{SM}}_{\gamma 1}\delta_{\gamma n}}{n!}\frac{\alpha}{\pi}\Big( \frac{h}{v} \Big)^{n} \nonumber + \cdots\\
& + c^{\text{SM}}_{\gamma Z1}\delta_{\gamma Z1}\frac{\alpha}{\pi v}h A_{\mu\nu} Z^{\mu\nu} +  A_{\mu\nu}Z^{\mu\nu} \sum_{n=2}^{\infty}\frac{c^{\text{SM}}_{\gamma Z1}\delta_{\gamma Zn}}{n!}\frac{\alpha}{\pi}\Big( \frac{h}{v} \Big)^{n} \nonumber + \cdots\\
& -\delta_{t1} \frac{m_{t}}{v}h \bar{t}t - \sum_{n=2}^{\infty}\frac{c_{tn}}{n!}\frac{m_{t}}{v}\bar{t}t + \nonumber \cdots
\end{flalign}
where $\delta_{x}$ parametrize the deviations in the couplings $g_{x}$ compared to the SM predictions\footnote{Notice that in the $\kappa$ framework, $\kappa_{x} = 1 + \delta_{x}$}
\begin{equation}\label{eq:delta}
\delta_{x} \equiv \frac{g_{x} - g_{x}^{(\text{SM})}}{g_{x}^{(\text{SM})}},
\end{equation}
whereas $c_{i}$ represent the Wilson coefficients of higher-dimensional operators that do not have SM counterparts, and $c^{\text{SM}}_{\gamma 1}$, $c^{\text{SM}}_{\gamma Z 1}$ are the SM effective Higgs coupling to $\gamma\gamma$ and $\gamma Z$ respectively, after integrating out the top and $W$ loops. Notice here that we have divided the operators by appropriate powers of $v$ to keep the Wilson coefficients dimensionless, i.e.  $v$ is NOT an expansion scale and is only introduced for convenience. Therefore, $\delta_{i}$ and $c_{i}$ could assume any value compatible with experimental measurements. It is more convenient to work in a gauge where the Goldstone-bosons are manifest so that we can use the equivalence theorem. We can do this first by writing the Higgs doublet as
\begin{equation}\label{eq:HiggsDoublet}
H = \frac{1}{\sqrt{2}} 
\begin{pmatrix}
G_{1} + i G_{2}\\
v + h + i G_{3}
\end{pmatrix},
\end{equation}
with $h$ being the physical Higgs field, and then defining the field
\begin{equation}\label{eq:Xfield}
X \equiv \sqrt{2H^{\dagger}H}-v = h+\frac{\vec{G}^{2}}{2(v+h)} - \frac{\vec{G}^{4}}{8(v+h)^{3}}+ O\Bigg( \frac{\vec{G}^{6}}{(v+h)^{5}}\Bigg),
\end{equation}
where we have introduced $\vec{G} = (G_{1},G_{2},G_{3})$\footnote{When calculating the amplitudes in this paper, we rewrite the Goldstone bosons in terms of the longitudinal modes of the $W$ and $Z$ using $W_{L}^{\pm} = \frac{1}{\sqrt{2}}(G_{1}\mp i G_{2})$ and $Z_{L} = G_{3}$.}. Because $X = h$ in the unitary gauge, we can generalize Eq. (\ref{eq:effLag1}) to any general gauge by substituting $h \rightarrow X$. 

Although Eq. (\ref{eq:effLag1}) appears similar to the Higgs Effective Field Theory (HEFT) \cite{Grinstein:2007iv}, we should nonetheless keep in mind that we are implicitly restoring the Higgs doublet via the above substitution. In fact, when Eq. (\ref{eq:effLag1}) is generalized to a general gauge using the aforementioned substitution, then one can show that it becomes isomorphic to SMEFT. More explicitly, the deviations $\delta_{i}$ and Wilson coefficients $c_{i}$ are assumed to receive contributions from multiple higher-order operators in the SMEFT expansion, such that one can match the generalized Lagrangian to the SMEFT expansion truncated at a certain order. To be more concrete, if we truncate SMEFT at a certain order, like dim-6 for example, and expand the operators explicitly; then one can map its parameters (i.e the cutoff scale and Wilson coefficients in SMEFT) to the parameters in our approach (i.e. $\delta_{i}$ and $c_{i}$). The correspondence is one-to-one for a certain SMEFT truncation but might change by including higher-order SMEFT operators. We show the matching explicitly in the $hgg$ sector in Subsection \ref{Sec:hggSMEFT}.

Operators in Eq. (\ref{eq:effLag1}) give rise to amplitudes that grow with energy, which eventually violate unitarity at some high energy scale, signaling the onset of new physics. To obtain the scale of new physics $E_{\text{max}}$, we demand that the amplitudes respect unitarity up to $E_{\text{max}}$, and only assume that there are no new light degrees of freedom up to that scale. We call processes that depend on one parameter only model-independent, as they do not depend on any assumptions regarding the possible UV completion. Other processes are not truly model-independent, because the scale of new physics can vary based on the assumed relations among the various parameters, which does depend on the particular UV completion.

For example, modification to the Higgs potential give rise to the energy-growing process $W_{L}^{+}W_{L}^{-}W_{L}^{+} \rightarrow W_{L}^{+}W_{L}^{-}W_{L}^{+}$. Demanding that the amplitude of this process respect unitarity gives the following model-independent bound on the scale of new physics
\begin{equation}\label{eq:H_Strongest1}
E_{\text{max}} \simeq \frac{14 \hspace{1mm}\text{TeV}}{|\delta_{3}|^{1/2}}.
\end{equation}

As clarified in \cite{Chang:2019vez}, loops contain either heavy particles or SM particles. In the former case, the heavy particles can be integrated out to give local terms that appear in the expansion in Eq. (\ref{eq:effLag1}). On the other hand, SM loops give perturbatively small corrections up to the unitarity-violating scale, where they become comparable to the tree-level contributions, and thus give $O(1)$ corrections.\footnote{In \cite{Abu-Ajamieh:2021vnh}, loops from Eq. (\ref{eq:effLag1}) were studied in detail to investigate the possibility of canceling the quadratic divergences in the Higgs mass corrections. There, it was found that the quadratic divergences in the Higgs mass could be canceled if the scale of new physics is $\lesssim 19$ TeV.} 

Finally, we should note that one could have additional operators in Eq. (\ref{eq:effLag1}) by introducing higher derivatives, however, higher derivatives will only lead to amplitudes that grow faster with energy and thus can only lower the unitarity-violating scale. We choose to be conservative and neglect operators with higher derivatives.

The technical details of how the states are defined and normalized, in addition to the derivation of the unitarity condition, are presented in more detail in \cite{Chang:2019vez, Abu-Ajamieh:2020yqi, Abu-Ajamieh:2021egq}.

\section{New Physics from the $gg$ Sector}
\label{Sec:hgg}
In this section, we extend the treatment in \cite{Chang:2019vez, Abu-Ajamieh:2020yqi, Abu-Ajamieh:2021egq} to the Higgs interaction with a pair of gluons. As the Higgs couples to $gg$ at loop-level, we would expect the scale of new physics to be larger than the tree-level ones found in \cite{Chang:2019vez, Abu-Ajamieh:2020yqi}, and comparable to the ones found in \cite{Abu-Ajamieh:2021egq} for the $\gamma\gamma$ and $\gamma Z$ couplings.

\subsection{Model-Independent Bound on the Scale of New Physics from the Coupling $hgg$}\label{Sec:hggBound1}
\begin{figure}[!t]
\centerline{\begin{minipage}{0.5\textwidth}
\centerline{\includegraphics[width=150pt]{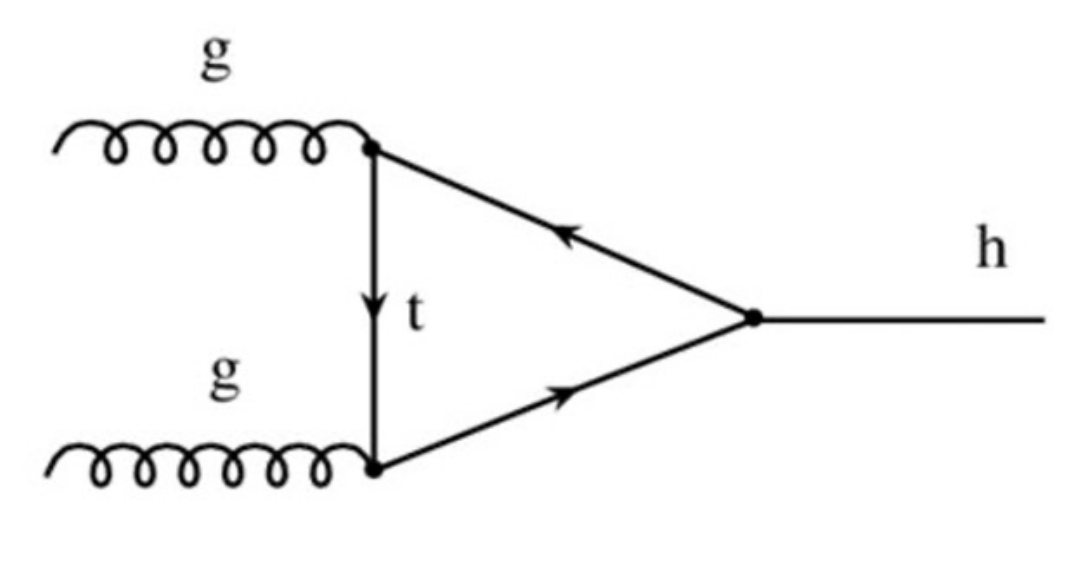}}
\caption{\small The SM coupling of the Higgs to $gg$. The contribution of the top quark is the dominant contribution.}
\label{fig1}
\end{minipage}}
\end{figure}

In the SM, the Higgs couples to a pair of gluons through a fermion loop as shown in Fig. \ref{fig1}. The contribution of the top quark dominates over all other fermions due to its larger Yukawa coupling, and one can integrate it out to write the coupling as a tree-level effective coupling. Using the convention in \cite{Carmi:2012in}, we can write the effective Lagrangian in the unitary gauge as
\begin{equation}\label{eq:Hgaga_Lag1}
\mathcal{L}_{hgg} = c_{g1} \frac{\alpha_{s}}{12\pi v} h G^{a}_{\mu\nu}G^{a\mu\nu},
\end{equation}
where $G^{a}_{\mu\nu} = \partial_{\mu}G^{a}_{\nu}-\partial_{\nu}G^{a}_{\mu}+ g f^{abc}G^{b}_{\mu}G^{c}_{\nu}$, and in the SM, $c^{\text{SM}}_{g 1} \simeq 1.03$ when only the top quark is retained. Similar to Eq. (\ref{eq:delta}), it is more convenient to express new physics in terms of the deviation in the $hgg$ coupling. Thus we define
\begin{equation}\label{eq:dev1}
\delta_{g 1} \equiv \frac{c_{g 1}-c^{\text{SM}}_{g 1}}{c^{\text{SM}}_{g 1}}.
\end{equation}

Moving away from the unitary gauge by restoring the Goldstones using Eq. (\ref{eq:Xfield}), and expressing the coupling $hgg$ in terms of $\delta_{g1}$, we write Eq. (\ref{eq:Hgaga_Lag1}) as
\begin{dmath}\label{eq:Hgaga_Lag2}
\mathcal{L}_{hgg} = \Big(\frac{\alpha_{s} c^{\text{SM} }_{g1} \delta_{g 1}}{12\pi v^{2}}\Big) \Big[Z_{L}^{2} + 2 W_{L}^{+}W_{L}^{-} \Big]\Big[(\partial_{\mu}G^{a}_{\nu})(\partial^{\mu}G^{a\nu})-(\partial_{\mu}G^{a}_{\nu})(\partial^{\nu}G^{a\mu})\Big]  +  \Big(\frac{\alpha_{s} c^{\text{SM} }_{g1} \delta_{g 1}}{12\pi v^{2}}\Big)g f^{abc} \Big[2 h v + Z_{L}^{2} + 2 W_{L}^{+}W_{L}^{-} \Big]\Big[ \partial_{\mu}G^{a}_{\nu} -\partial_{\nu}G^{a}_{\mu} \Big]G^{b\mu}G^{c\nu} + \Big( \frac{\alpha_{s}^{2}c^{\text{SM}}_{g1}\delta_{g1}}{6v^{2}}\Big)(f^{abc})^{2}\Big[2 h v + Z_{L}^{2} + 2 W_{L}^{+}W_{L}^{-} \Big] (G^{b}_{\mu}G^{b\mu})(G^{c}_{\nu}G^{c\nu}) + \cdots,
\end{dmath}
where the ellipsis indicates higher-order operators that depend on other deviations and/ or Wilson coefficients. In this paper, we limit ourselves to the contact interactions and neglect processes with propagators.\footnote{For a complete discussion of processes with propagators and the possible IR enhancement, see Appendix A5 of \cite{Abu-Ajamieh:2020yqi}.} 

Before we discuss the bounds, we need to mention a few comments on Eq. (\ref{eq:Hgaga_Lag2}): As we can see from the above equation, due to the non-Abelian covariant derivative, we have three types of operator. The first type contains two derivatives, the second contains one, and the third contains no derivatives. Inspecting the operators that contain only one derivative, it is easy to see that they give vanishing contributions, as first and second terms give identical contributions that cancel one another. This result is general: All operators with one derivative have vanishing amplitudes. As for the remaining two types, notice that the amplitudes of the operators with two derivatives will have an enhanced energy growth compared to the ones without derivatives. This in general will make their amplitudes grow faster and thus violate unitarity at lower scales, resulting in generally stronger bounds.\footnote{In calculating processes with 4 gluons, we simplify our calculation by summing and averaging over initial and final spins, whereas we do not do so for processes with 2 gluons only. Our results are not impacted by this.}

Extracting the unitarity bounds from Eq. (\ref{eq:Hgaga_Lag2}), we find that the strongest bound comes from $g_{\pm} g_{\pm} \rightarrow W_{L}^{+}W_{L}^{-}$ and reads
\begin{equation}\label{eq:delta1_strongest}
E_{\text{max}} \simeq \frac{10.9 \hspace{1mm} \text{TeV}}{|\delta_{g1}|^{1/2}}.
\end{equation}

\begin{figure}[!t]
\centerline{\begin{minipage}{0.8\textwidth}
\centerline{\includegraphics[width=300pt]{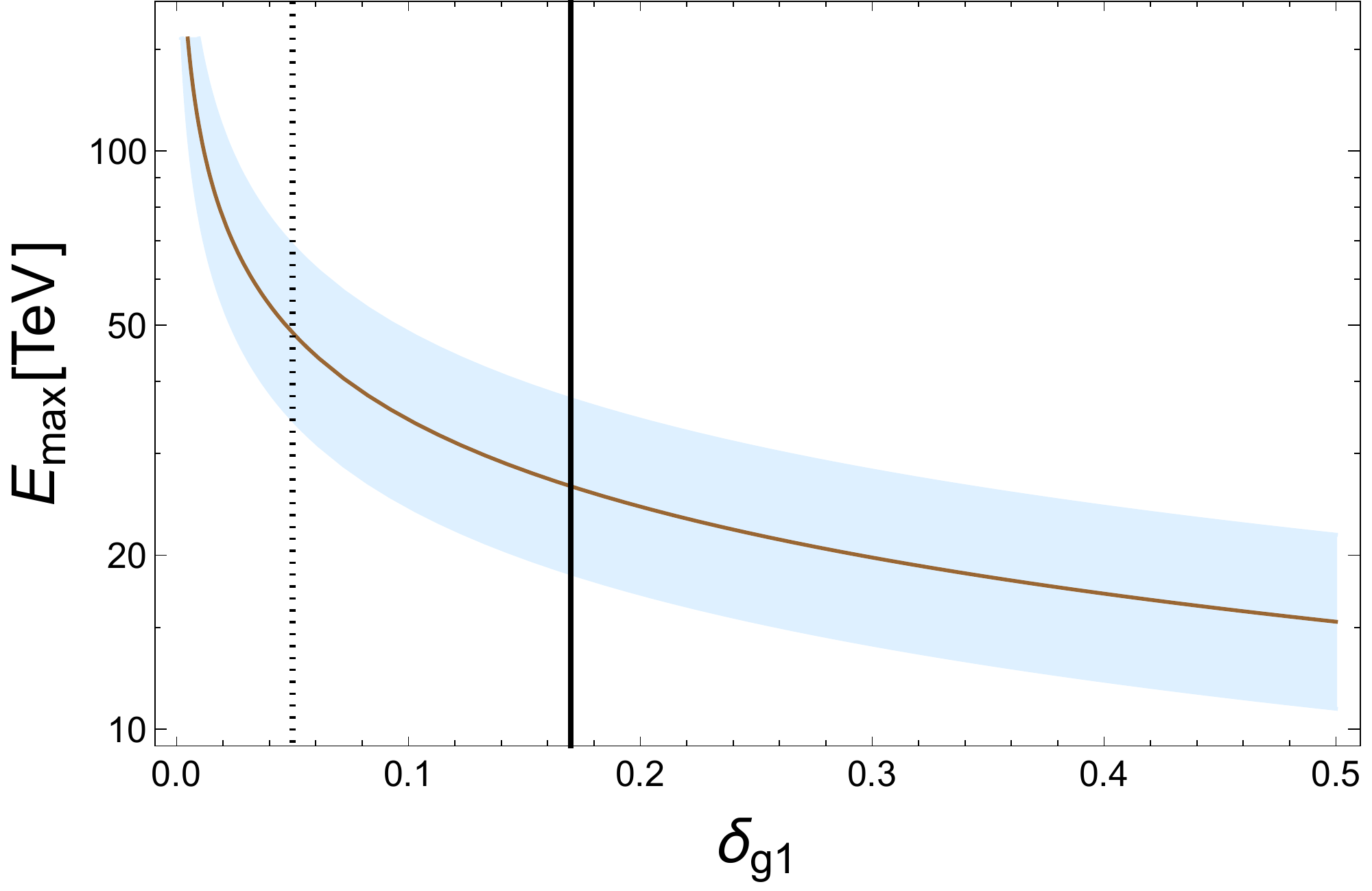}}
\caption{\small 
The unitarity-violating bound as a function of $\delta_{g 1}$. The model-independent bound is obtained from Eq. (\ref{eq:delta1_strongest}). The band around the model-independent bound results from varying the unitarity bound to $\frac{1}{2} \leq \hat{\mathscr{M}} \leq 2$. The solid black line represents the $95\%$ C.L. limits on $\delta_{g 1}$ obtained from ATLAS \cite{ATLAS:2019nkf}, whereas the dotted one represents the $95\%$ C.L. HL-LHC projections obtained from \cite{Cepeda:2019klc}.}
\label{fig2}
\end{minipage}}
\end{figure}

The $95\%$ limits from ATLAS \cite{ATLAS:2019nkf} constrain $\delta_{g 1} \in [-0.11,0.17]$, which gives a scale of new physics $\gtrsim 26$ TeV. This is beyond the reach of the LHC, however, it is well within the reach of the future $100$ TeV collider. In addition, $\delta_{g 1}$ can be probed indirectly in the HL-LHC, which will help make the limits tighter. For instance, \cite{Cepeda:2019klc} puts the HL-LHC's $95\%$ projections for $\delta_{g 1}$ at $ \pm 5\%$, which would be push the scale of new physics to above $\sim 48$ TeV.  We plot the unitarity bound as a function of $\delta_{g 1}$ in Figure \ref{fig2}, together with the LHC limits and HL-LHC projections.

\subsection{Model-Independent Bounds on the Scale of New Physics from the Couplings $h^{n}gg$}\label{Sec:hggBoundn}
\begin{figure}[!t]
\centerline{\begin{minipage}{\textwidth}
\centerline{\includegraphics[width=350pt]{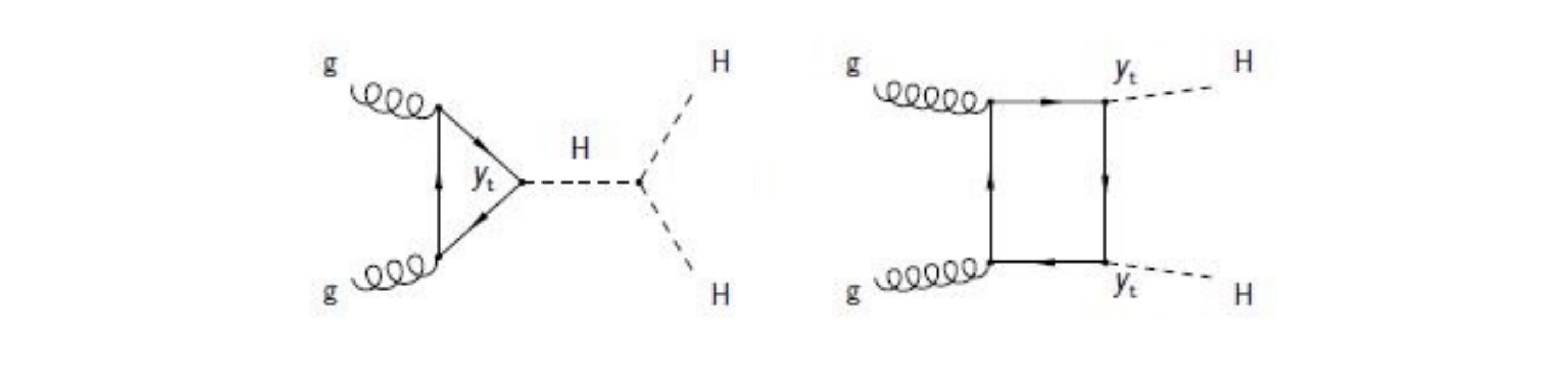}}
\caption{\small The SM coupling of two Higgses to $gg$.}
\label{fig4}
\end{minipage}}
\end{figure}

Eq. (\ref{eq:Hgaga_Lag1}) can easily be generalized to include interactions with any number of Higgs bosons with a gluon pair by writing the effective Lagrangian as
\begin{equation}\label{eq:nHgaga_Lag1}
\mathcal{L}_{h^{n}gg} = G^{a}_{\mu\nu}G^{a\mu\nu} \sum_{n=1}^{\infty}\frac{c_{g n}}{n!}\frac{\alpha_{s}}{12\pi}\Big( \frac{h}{v} \Big)^{n},
\end{equation}
where $c_{g n}$ are Wilson coefficients that parametrize the effective couplings of $n$-Higgses with a gluon pair. We point out that $c_{g n}$ do not vanish in the SM. For example, $c^{\text{SM}}_{g 2}$ is obtained by integrating the propagator and loops in the diagrams shown in Figure  \ref{fig4}. However, unlike $c^{\text{SM}}_{g 1}$, $c^{\text{SM}}_{g n}$ for $n \geq 2$ have never been calculated to the best of our knowledge. Thus, it is more convenient to define the deviations $\delta_{g n}$ in terms of $c^{\text{SM}}_{g 1}$ instead
\begin{equation}\label{eq:delta_ga_n}
\delta_{g n} \equiv \frac{c_{g n}-c^{\text{SM}}_{g n}}{c^{\text{SM}}_{g 1}}, \hspace{5mm} \text{for} \hspace{2mm} n \geq 2,
\end{equation}
and treat $\delta_{g n}$ as free parameters. Notice here that defining $\delta_{g n}$ in this manner does not quantify how far we deviate from the SM predictions precisely, however, it does allow us to express the SM limit more transparently as $\delta_{g n} \rightarrow 0$. We can restore the Goldstone bosons in the usual way and then expand the fields in Eq. (\ref{eq:nHgaga_Lag1})
\begin{dmath}\label{eq:nHgaga_Lag2}
\mathcal{L}_{h^{n}gg} = \frac{\alpha_{s} c^{\text{SM} }_{g1}}{24\pi} G^{a}_{\mu\nu}G^{a\mu\nu} \Big\{ \Big(\frac{\delta_{g1}}{v^{2}} \Big) \Big[Z_{L}^{2} + 2 W_{L}^{+}W_{L}^{-} \Big] + \Big(\frac{\delta_{g2}}{v^{2}} \Big)h^{2} + 
 \Big( \frac{\delta_{g 2}-\delta_{g 1}}{v^{3}}\Big)h \Big[Z_{L}^{2} + 2 W_{L}^{+}W_{L}^{-} \Big]  + \Big(\frac{\delta_{g 3}}{3v^{3}} \Big)h^{3} + 
 \Big( \frac{\delta_{g2}- \delta_{g 1}}{4v^{4}}\Big) \Big[Z_{L}^{2} + 2 W_{L}^{+}W_{L}^{-} \Big]^{2}+ 
 \Big( \frac{\delta_{g1}-\delta_{g2}+\frac{1}{2}\delta_{g3}}{v^{4}}\Big) h^{2} \Big[  Z_{L}^{2} + 2 W_{L}^{+}W_{L}^{-}\Big] + \Big(\frac{\delta_{g 4}}{12v^{4}} \Big)h^{4} + \dots 
\Big\} + \frac{\alpha_{s} c^{\text{SM}}_{1}}{6\pi}g f^{abc}G^{a}_{\mu\nu}G^{b\mu}G^{c\nu} \Big\{\frac{\delta_{g1}}{v}h + \frac{\delta_{g1}}{2v^{2}}\Big[Z_{L}^{2} + 2 W_{L}^{+}W_{L}^{-} \Big] + \frac{\delta_{g2}}{2v^{2}}h^{2} + \Big( \frac{\delta_{g 2}-\delta_{g 1}}{2v^{3}}\Big)h \Big[Z_{L}^{2} + 2 W_{L}^{+}W_{L}^{-} \Big] + \frac{\delta_{3}}{6v^{3}}h^{3} +\cdots \Big \} + \frac{\alpha_{s}^{2}c^{\text{SM}}_{g1}}{3}(f^{abc})^{2}( G^{b}_{\mu}G^{b\mu})(G^{c}_{\nu}G^{c\nu}) \Big\{ \frac{\delta_{g1}}{v}h + \frac{\delta_{g1}}{2v^{2}}\Big[Z_{L}^{2} + 2 W_{L}^{+}W_{L}^{-} \Big] + \frac{\delta_{g2}}{2v^{2}}h^{2} + \cdots \Big\}.
\end{dmath}

Notice here that Eq. (\ref{eq:Hgaga_Lag2}) can be obtained from Eq. (\ref{eq:nHgaga_Lag2}) by setting $\delta_{g n} \rightarrow 0$ for $n > 1$. Also notice that as mentioned above, the second term which contains a single derivative gives vanishing amplitudes, and the last term which has no derivatives gives subleading bounds in general, although they might dominate for scatterings with larger particles as we shall see below. Up to the dim-8 operators shown above, the strongest model-independent bounds are given by the following processes

\begin{dmath}\label{eq:nHgaga_Strongest}
\begin{aligned}
g_{\pm}g_{\pm} \rightarrow hh & :  \hspace{2mm} E_{\text{max}} \simeq \frac{12.9 \hspace{1mm}\text{TeV}}{|\delta_{g 2}|^{1/2}}, \\
g_{\pm}g_{\pm} \rightarrow hhh & :  \hspace{2mm} E_{\text{max}} \simeq \frac{10.8 \hspace{1mm}\text{TeV}}{|\delta_{g 3}|^{1/3}}, \\
g_{\pm}g_{\pm}h \rightarrow hhh & : \hspace{2mm} E_{\text{max}} \simeq \frac{11.3 \hspace{1mm}\text{TeV}}{|\delta_{g 4}|^{1/4}}. \\
\end{aligned}
\end{dmath}

Given that $\delta_{g n}$ are essentially free parameters, Eq. (\ref{eq:nHgaga_Strongest}) seems to suggest that the scale of new physics could be significantly low and perhaps even within the reach of the LHC. However, this would require $\delta_{g n} \gtrsim 2$ to have any hope of being probed in the LHC, which is highly unlikely. Nonetheless, the scale of new physics from these processes is relatively low and is well within the reach of the $100$ TeV collider, even for small values of the deviations. For example, for $\delta_{g2,3,4} \sim 0.1$, the scale of new physics could be between $\sim 20 - 40$ TeV. In addition, the HL-LHC could indirectly probe these deviations and thus set lower bounds on the scale of new physics.

The processes given in Eq. (\ref{eq:nHgaga_Strongest}) are not the only model-independent ones that can be obtained from the Lagrangian in Eq. (\ref{eq:nHgaga_Lag2}). Inspecting Eq. (\ref{eq:nHgaga_Lag2}), it is not hard to see that all processes involving $g^{2}$ or $g^{4}$ with any number of pure Higgses are indeed model-independent. To be more specific, we consider the following type of operators
\begin{equation}\label{eq:2n+2_model_indep}
\delta \mathcal{L}_{\text{2n+2}} = \Big( \frac{c_{g 1}^{\text{SM}}\delta_{g2n}\alpha_{s}}{12\pi(2n)!}\Big) \Big( \frac{h}{v}\Big)^{2n} G^{a}_{\mu\nu}G^{a\mu\nu} + \Big( \frac{c_{g 1}^{\text{SM}}\delta_{g2n-2}\alpha_{s}^{2}}{3(2n-2)!}\Big)(f^{abc})^{2} \Big(\frac{h}{v}\Big)^{2n-2} (G_{\mu}^{b}G^{b\mu})(G_{\nu}^{c}G^{c\nu}).
\end{equation}

\begin{figure}[!t]
\centerline{\begin{minipage}{0.8\textwidth}
\centerline{\includegraphics[width=300pt]{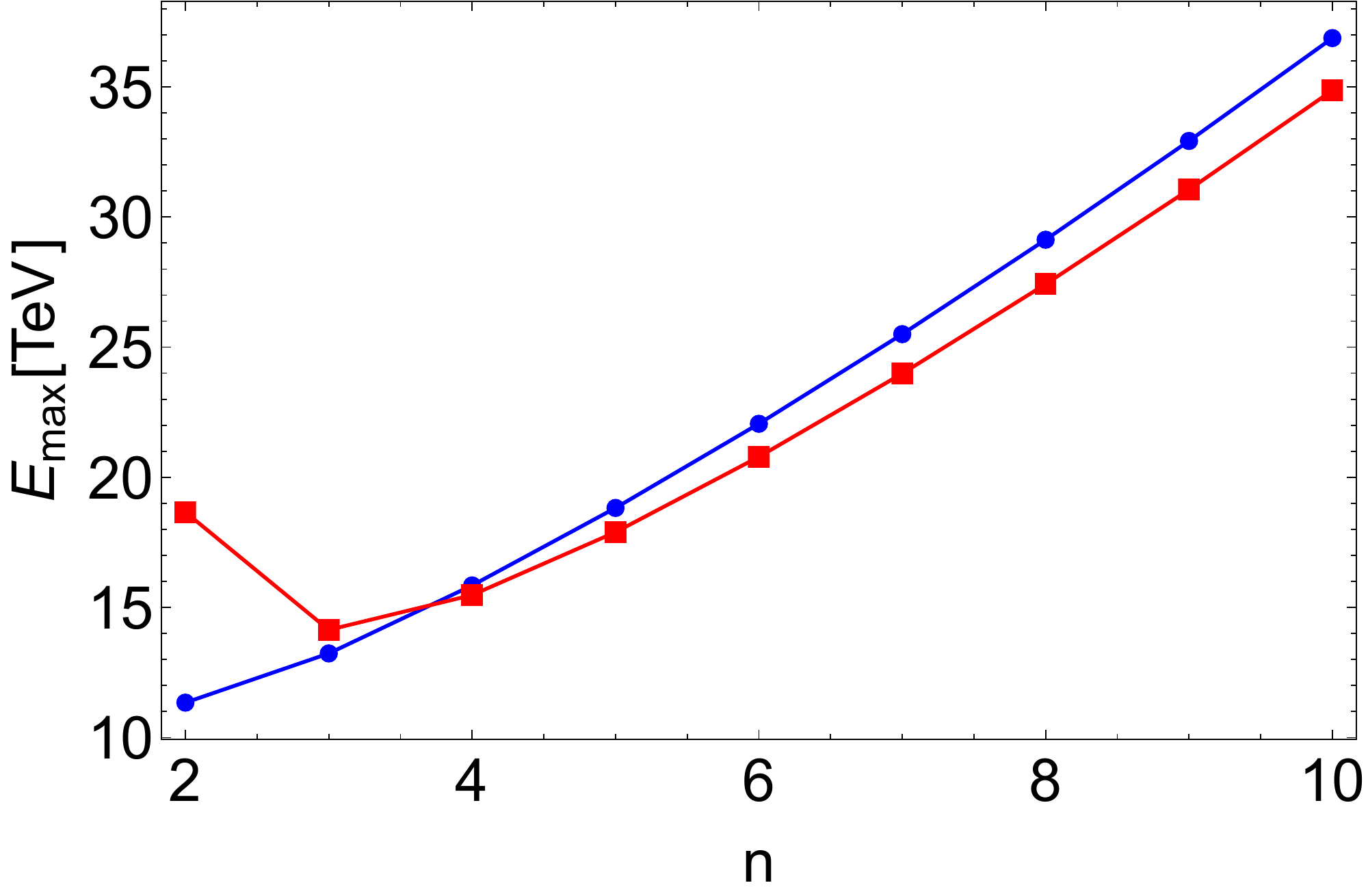}}
\caption{\small 
The unitarity-violating scale as a function of $n$ from $g_{\pm}g_{\pm} h^{n-1}\rightarrow h^{n+1}$ (blue), and $ggh^{n-1} \rightarrow ggh^{n-1} $ (red). Both processes are comparable for $n \geq 3$. Here $\delta_{g 2n}, \delta_{g 2n-2}$ are set to $1$.}
\label{fig5}
\end{minipage}}
\end{figure}

The strongest bounds are obtained from symmetric processes, i.e. processes with an equal number of initial and final states. This fact stems purely from combinatorics as shown in detail in \cite{Chang:2019vez}. We find that the following processes are the dominant ones
\begin{dmath}\label{eq:2n_bounds}
\begin{aligned}
g_{\pm}g_{\pm} h^{n-1}\rightarrow h^{n+1} & :  \hspace{2mm} E_{\text{max}} = 4\pi v \Bigg( \frac{3(n+1)!n!\sqrt{(n+1)!(n-1)!}}{\sqrt{2N_{c}} \alpha_{s} c^{\text{SM}}_{g 1}|\delta_{g 2n}|}\Bigg)^{\frac{1}{2n}}, \\
ggh^{n-1}\rightarrow ggh^{n-1} & : \hspace{2mm}  E_{\text{max}} = 4\pi v \Bigg(\frac{2\pi (n-1)!(n-1)!n!}{N_{c}\alpha_{s}^{2} C(G) c^{\text{SM}}_{g 1} |\delta_{g 2n-2}|}\Bigg)^{\frac{1}{2n-2}}. \\
\end{aligned}
\end{dmath}
where $N_{c} = 8$ is a color factor, $C(G)$ is the Casimir invariant and $n \geq 2$. Figure \ref{fig5} shows a comparison between the two processes in Eq. (\ref{eq:2n_bounds}), where we set $\delta_{g2n},\delta_{g2n-2} \rightarrow 1$. From the figure, we can see that $ggh^{n-1} \rightarrow ggh^{n-1}$ dominates over $g_{\pm}g_{\pm}h^{n-1} \rightarrow h^{n+1}$ for $n \geq 3$. This is counter-intuitive, as the latter process has a stronger energy dependence by two powers as can be seen from Eq. (\ref{eq:2n_bounds}), and although we have set both $\delta_{g2n}$ and $\delta_{g2n-2}$ to $1$, which needn't be the case, we have checked this is behavior persists as long as $\delta_{g2n} \leq \delta_{g2n-2}$, which is quite a plausible assumption. The main reason for this is that $ggh^{n-1} \rightarrow ggh^{n-1}$ is more symmetric than the other process, and thus the enhacement due to the combinatorics more than compenstates for the extra two derivatives, especially at large $n$ where the $(2n)^{\text{th}}$ and $(2n-2)^{\text{th}}$ roots are not significantly different. 

We can also see that the strongest bound comes from $n \sim 2-4$, which seems to suggest that the bounds given in Eq. (\ref{eq:nHgaga_Strongest}) are not far from optimal. However, as in Figure \ref{fig5} we have set $\delta_{g2n} = \delta_{g2n-2} =1$, it is possible in principle that different values might change this behavior. However, we checked that this behavior is maintained as long as these deviations become smaller with $n$, which, as we stated, is quite conservative and consistent with an EFT expansion. We have also checked that the minimum is not too sensitive to specific values of $\delta_{g 2n}$ and $\delta_{g2n-2}$ and remains around $\sim 2-4$. Thus, we conclude that the strongest unitarity-violating scale should not be too far from the ones given in Eq. (\ref{eq:nHgaga_Strongest}).

\subsection{Other Bounds}\label{Sec:hggMixedProcesses} 
The processes given in Eqs. (\ref{eq:delta1_strongest}), (\ref{eq:nHgaga_Strongest}) and (\ref{eq:2n_bounds}) are the only model-independent ones in this sector. Nonetheless, it is still worthwhile investigating other processes which are not model-independent. The reason for this is that for a certain choice of the NLO deviations, the scale of new physics could be lower than the model-independent ones, and could potentially be probed in the HL-LHC. Furthermore, studying such processes can help constrain higher-order deviations that are usually harder to measure directly. Here, we limit ourselves to the unitarity-violating processes that depend only on $\delta_{g 1}$ and $\delta_{g 2}$. Using the general Lagrangian in Eq. (\ref{eq:nHgaga_Lag2}), the strongest bounds are 
\begin{dmath}\label{eq:NLObounds}
\begin{aligned}
g_{\pm}g_{\pm} \rightarrow W_{L}^{+}W_{L}^{-} h & : \hspace{2mm} E_{\text{max}} \simeq \frac{8 \hspace{1mm}\text{TeV}}{|\delta_{g 2} - \delta_{g 1}|^{1/3}}, \\
g_{\pm}g_{\pm} W_{L}^{+} \rightarrow W_{L}^{+}W_{L}^{-}W_{L}^{+}  & :  \hspace{2mm} E_{\text{max}} \simeq \frac{8.3 \hspace{1mm}\text{TeV}}{|\delta_{g 2} - \delta_{g 1}|^{1/4}}. \\
\end{aligned}
\end{dmath}

\begin{figure}[!t]
\centerline{\begin{minipage}{0.8\textwidth}
\centerline{\includegraphics[width=300pt]{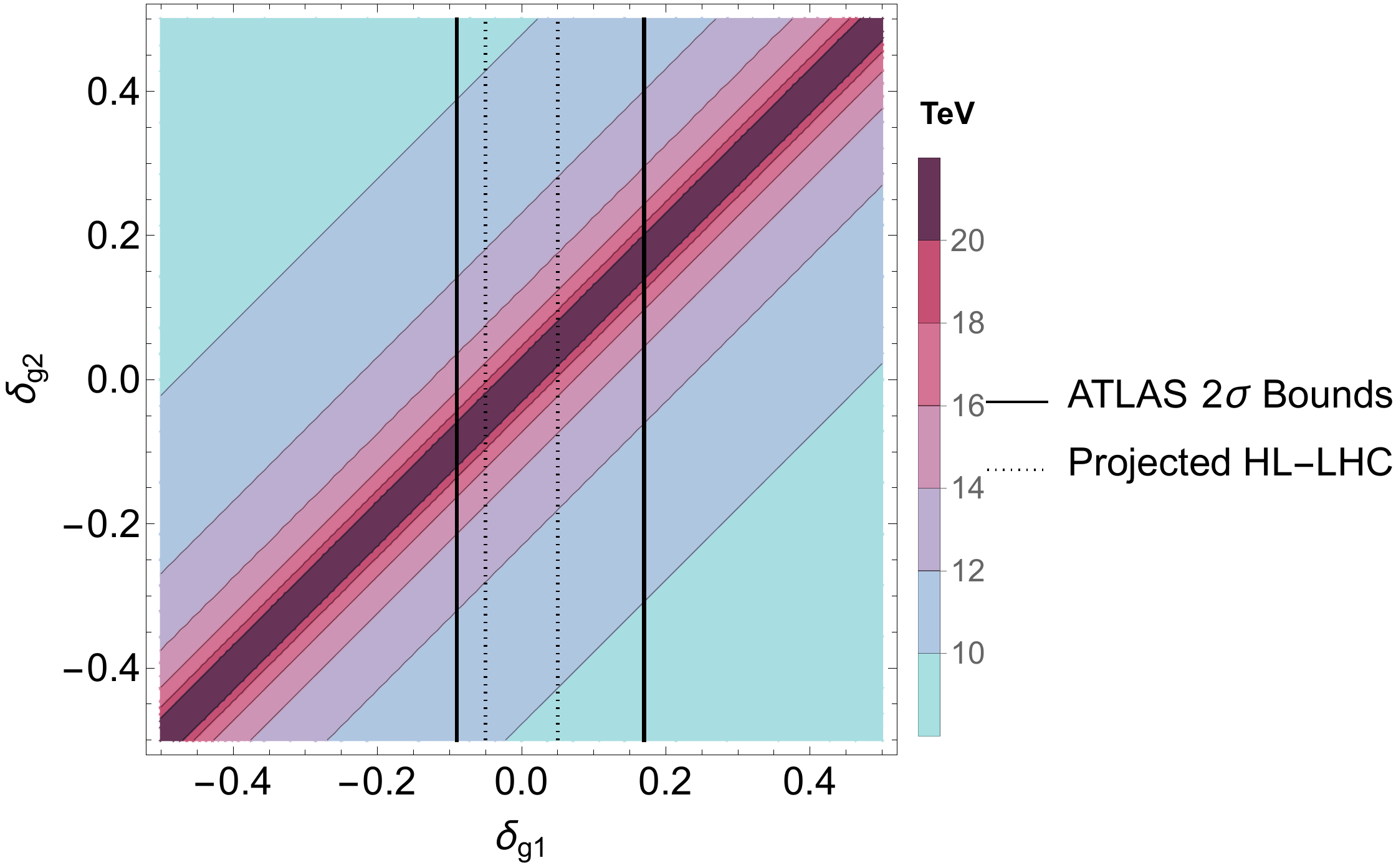}}
\caption{\small 
The unitarity-violating scale that depends on both $\delta_{g1}$ and $\delta_{g2}$. The solid black line represents the $95\%$ C.L. limits on $\delta_{g 1}$ from ATLAS \cite{ATLAS:2019nkf}, whereas the dotted line represents the $95\%$ HL-LHC projections \cite{Cepeda:2019klc}. Here we set an upper limit on $|\delta_{g1,2}| \hspace{1mm}\leq 0.5$.}
\label{fig6}
\end{minipage}}
\end{figure}

We show the strongest unitarity-violating scale from Eq. (\ref{eq:NLObounds}), in Figure \ref{fig6}, together with the ATLAS  limits \cite{ATLAS:2019nkf} and HL-LHC projections \cite{Cepeda:2019klc} superimposed. Notice that the diagonal, which corresponds to $\delta_{g 2} =\delta_{g 1}$, gives a unitarity-violating scale that blows up, as one can evidently observe from Eq. (\ref{eq:NLObounds}). The interpretation of this behavior is very simple: It reflects the fact that the corresponding operators in (\ref{eq:nHgaga_Lag2}) vanish, and thus one needs to include the NLO operator. The plot also reveals that the scale of new physics could potentially be smaller than the model-independent ones. For example, if we make the conservative assumption that $|\delta_{g 2}| \sim 0.4$, then the scale of new physics could be as low as $\sim 10$ TeV. Which, although is beyond the reach of the LHC, is more accessible in future colliders such as the $100$ TeV collider. In addition, one can use the limits on $\delta_{g 1}$ to set bound $\delta_{g 2}$ using any null collider searches.
\subsection{SMEFT Predictions from Unitarity}
\label{Sec:hggSMEFT}
As mentioned in the introduction, the SMEFT approach is predicated upon the assumption that the UV theory is of the decoupling type, which means that in the low-energy theory, the effects of new physics should be captured by adding to the SM higher-dimensional gauge-invariant operators that become more and more irrelevant the higher the scale of new physics becomes. As we mentioned before, SMEFT does not provide a way to estimate the uncertainty associated with neglecting higher-order operators. It was shown in \cite{Abu-Ajamieh:2020yqi} that using unitarity arguments alone could make a quantitative statement about the accuracy of SMEFT.

To this avail, notice that for the SMEFT prescription to be plausible, then the leading operators in the expansion, i.e. the dim-6 operators, need to dominate over the next-to-leading ones. To be more concrete, we consider a theory consisting of the SM in addition to the dim-6 SMEFT operator
\begin{equation}\label{eq:ga_dim6_SMEFT}
\mathcal{L}^{\text{dim-6}}_{\text{SMEFT}} = \frac{1}{M^{2}}\Big( H^{\dagger}H - \frac{v^{2}}{2}\Big) G^{a}_{\mu\nu}G^{a\mu\nu},
\end{equation}
and demand that it dominate over all higher-order operators. From this, we find that SMEFT predicts the following relations
\begin{equation}\label{eq:ga_SMEFTpred}
\delta_{g 1} = \frac{12\pi v^{2}}{\alpha_{s} c^{\text{SM}}_{g1}M^{2}}, \hspace{10mm} \delta_{g 2} = \delta_{g 1}, \hspace{10mm} \delta_{g n} = 0 \hspace{2 mm} \text{for} \hspace{2mm} n  \geq 3.
\end{equation}

\begin{figure}[!t]
\centerline{\begin{minipage}{0.8\textwidth}
\centerline{\includegraphics[width=300pt]{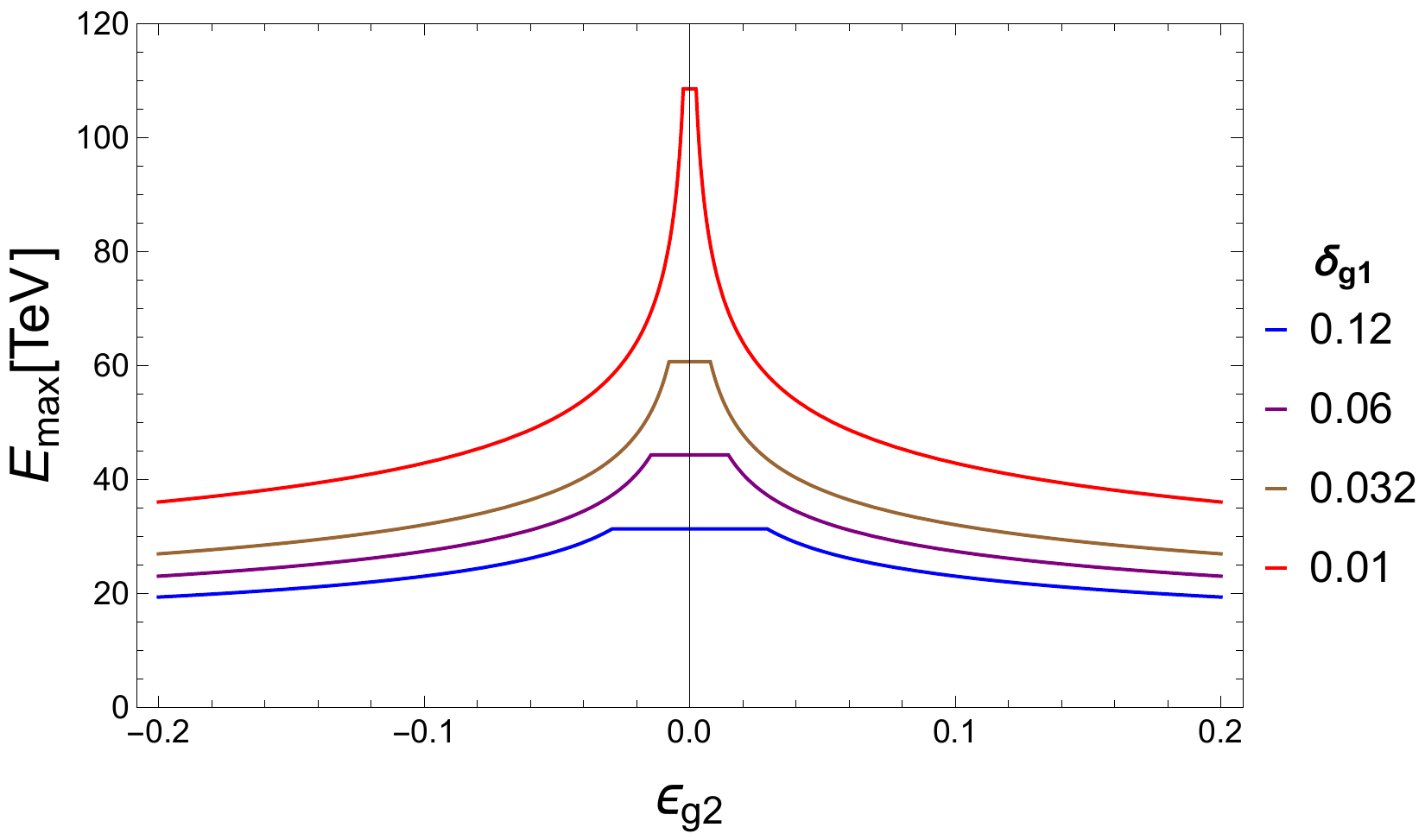}}
\caption{\small 
The unitarity-violating scale from processes that depend on both $\delta_{g 1}$ and $\delta_{g 2}$ as function of the fractional deviation $\epsilon_{g 2}$ from the predictions of the dim-6 SMEFT operator. See Eqs. (\ref{eq:ga_SMEFTpred}) and (\ref{eq:ga_epsilon}).}
\label{fig7}
\end{minipage}}
\end{figure}

We can estimate the accuracy of SMEFT by requiring that the scale of new physics that comes from $\delta_{g 2}$ does not fall below that coming from  $\delta_{g 1}$. This requirement translating into tuning $\delta_{g 2}$ to be close to the SMEFT prediction in Eq. (\ref{eq:ga_SMEFTpred}), as can be easily seen by inspecting Eq. (\ref{eq:NLObounds}), where we can see that the unitarity-violating scale becomes larger when $\delta_{g 2}$ is tuned to be close to $\delta_{g 1}$. To be more quantitative, we parametrize the deviation in $\delta_{g 2}$ from the SMEFT prediction as
\begin{equation}\label{eq:ga_epsilon}
\epsilon_{g 2} \equiv \frac{\delta_{g 2} - \delta^{\text{SMEFT}}_{g 2}}{\delta^{\text{SMEFT}}_{g 2}} \ll 1,
\end{equation}
where $\delta_{g 2}^{\text{SMEFT}}$ is given by Eq. (\ref{eq:ga_SMEFTpred}). The process that is most sensitive to new physics is given by the second line in Eq. (\ref{eq:NLObounds}). We plot the bound from this process against the fractional deviation from the SMEFT predictions $\epsilon_{g 2}$ in Figure \ref{fig7}. The plot reveals that SMEFT becomes more accurate the higher the scale of new physics is, which is what one expects for an EFT description. For example, Suppose the HL-LHC measures $|\delta_{g 1}|$ to be within $2.5 \%$ at $1$-$\sigma$ as projected in \cite{Cepeda:2019klc}, then Eq. (\ref{eq:delta1_strongest}) would predict a scale of new physics around $\sim 69$ TeV, which in turn would limit $|\delta_{g 2}|$ to be less than $\lesssim 1\%$ of the SMEFT predictions as can be seen from Figure \ref{fig7}. Thus, not only do precision measurements of $\delta_{g 1}$ predict the scale of new physics, but they also can help constrain $\delta_{g 2}$, which is hard to measure directly.

\section{Possible UV Completions}\label{sec:UVcompletion}
In this section, we present a few simple extensions to the SM and show how our EFT approach is mapped to the UV completion. In particular, we consider the 2-Higgs Doublet Model (2HDM) - Type II, the scalar singlet extension of the Higgs sector, and the extension of the SM with a top quark partner. In each scenario, we briefly summarize the UV model, show how the leading deviations map into the UV parameters and calculate the unitarity-violating scale given the experimental constraints on the model's parameters.
\subsection{The 2HDM - Type II}
In the 2HDM, the Higgs sector is extended to 2 doublets: an up-type doublet and a down-type doublet, each developing its own VEV
\begin{equation}\label{eq:2HDM}
H_{u} = \begin{pmatrix}
H^{+}_{u}\\
\frac{v_{u}+H^{0}_{u}}{\sqrt{2}}
\end{pmatrix}, \hspace{1cm}
H_{d} = \begin{pmatrix}
\frac{v_{d}+H^{0}_{d}}{\sqrt{2}}\\
H^{-}_{d}\\
\end{pmatrix}.
\end{equation}
where $\tan{\beta} \equiv v_{u}/v_{d}$, $\frac{\pi}{2} > \beta > 0$. The 2HDM has 8 degrees of freedom, 3 of which are eaten by the $W^{\pm}$ and $Z$, whereas the rest give rise to 2 $CP$-even neutral Higgses $h$ and $H$, 1 $CP$-odd neutral Higgs $A$, and two charged Higges $H^{\pm}$. The physical $CP$-even neutral Higgses are obtain via the rotation
\begin{equation}\label{eq:2HDMRot}
\begin{pmatrix} 
H\\
h
\end{pmatrix} = 
\begin{pmatrix} 
\cos{\alpha} && \sin{\alpha}\\
-\sin{\alpha} && \cos{\alpha}
\end{pmatrix} 
\begin{pmatrix} 
\text{Re}(H^{0}_{d})\\
\text{Re}(H^{0}_{u})
\end{pmatrix}. 
\end{equation}
where $h$ is the SM Higgs. In type II, $H_{u}$ couples to the up-type quarks, whereas $H_{d}$ couples to down-type quarks and leptons. This implies that the coupling of $h$ to the top quark gets multiplied by $\frac{\cos{\alpha}}{\sin{\beta}}$. Therefore, the value of $c^{\text{2HDM}}_{g1}$ which result from integrating out the top loop is obtained by simply modifying the SM contributions accordingly, and $\delta_{g1}$ is obtained from Eq. (\ref{eq:dev1}). Given the bounds on $\alpha$ and $\beta$ in the 2HDM - Type II \cite{Atkinson:2021eox}, the unitarity-violating scale is found to be
\begin{eqnarray}\label{eq:2HDM_Emax}
E^{\text{max}}_{gg} \gtrsim 34.5 \hspace{1mm} \text{TeV}.
\end{eqnarray}
\subsection{A Scalar Singlet Extension}
We consider extending the SM Higgs sector with a scalar singlet $\Phi$ that develops a VEV $v_{1}$. In the unitary gauge, the SM scalar sector becomes
\begin{equation}\label{eq:SMH}
H = \begin{pmatrix}
0\\
\frac{v_{0}+\phi}{\sqrt{2}}
\end{pmatrix}, \hspace{1cm}
\Phi = v_{1} + \chi,
\end{equation}
where $v_{0}$ is the SM Higgs VEV and $\phi$ and $\chi$ are the gauge eigenstates of the SM Higgs and scalar singlet respectively. The part of the Lagrangian that is relevant for our study is given by
\begin{equation}\label{eq:Singlet_Lag}
\mathcal{L} = (D_{\mu}H)^{\dagger}(D^{\mu}H) + \frac{1}{2}(\partial_{\mu} \Phi)(\partial^{\mu}\Phi) + V(H,\Phi) -\big( y_{t} \bar{t}_{L}\tilde{H}t_{R} + h.c.\big),
\end{equation}
where $\tilde{H} = \epsilon H^{*}$, $y_{t}$ is the SM top quark Yukawa coupling, and the scalar potential is given by
\begin{equation}\label{eq:ScalarV}
V(H,\Phi) = m_{H}^{2}(H^{\dagger}H) +\frac{1}{2} m_{\Phi}^{2}\Phi^{2} + \lambda (H^{\dagger}H)^{2} - \mu \Phi (H^{\dagger}H).
\end{equation}

The potential in Eq. (\ref{eq:ScalarV}) is minimized by imposing $\partial V/\partial v_{0} = \partial V/\partial v_{1} = 0$, and the mass matrix is obtained by substituting Eq. (\ref{eq:SMH}) in $V(H,\Phi)$. The masses of the physical SM Higgs $h$ and heavy Higgs $S$ are obtained through diagonalizing the mass matrix, which gives the following mass eigenvalues for the SM Higgs $h$ and the heavy Higgs $S$
\begin{equation}\label{eq:HiggsMasses}
M_{h,S}^{2} = \frac{1}{2}m_{\Phi}^{2} + \lambda v_{0}^{2} \mp \frac{1}{2} \sqrt{(m_{\Phi}^{2}-2\lambda v_{0})^{2}+4\mu^{2}v_{0}^{2}}.
\end{equation}

Notice that in the limit $\mu \rightarrow 0$, the heavy Higgs $S$ decouples and the SM limit is restored. The mass eigenstates $h$ and $S$ can be obtained from the gauge eigenstates $\phi$ and $\chi$ via the rotation
\begin{equation}\label{eq:SingletRot}
\begin{pmatrix} 
h\\
S
\end{pmatrix} = 
\begin{pmatrix} 
\cos{\alpha} && \sin{\alpha}\\
-\sin{\alpha} && \cos{\alpha}
\end{pmatrix} 
\begin{pmatrix} 
\phi\\
\chi
\end{pmatrix},
\end{equation}
where
\begin{eqnarray}\label{eq:alpha_Singlet}
\sin{2\alpha} = \frac{2 \mu v_{0}}{M_{S}^{2}-M_{h}^{2}},\\
\tan{\alpha} = \frac{\mu v_{0}}{M_{h}^{2}-2\lambda v_{0}^{2}}.
\end{eqnarray}

A straightforward calculation shows that the coupling of the SM Higgs to the top quark gets modified by $\cos{\alpha}$. Therefore, $c_{g1}$ in this scenario gets simply multiplied by $\cos{\alpha}$, and the deviation $\delta_{g1} = \cos{\alpha} -1$. The constraints on $\alpha$ are obtained from the Higgs signal strength \cite{ATLAS:2019nkf}
\begin{equation}\label{eq:SingletBound}
\cos^{2}{(\alpha)} \in [0.95,1].
\end{equation}
and with this bound, the unitarity-violating scale in this scenario is found to be
\begin{eqnarray}\label{eq:Singlet_Emax}
E^{\text{max}}_{gg} \gtrsim 68 \hspace{1mm} \text{TeV}.
\end{eqnarray}

\subsection{Sinlge Top Partner} 
Finally, we consider extending the SM with a single top quark partner $T$, that couples to the SM Higgs via the interaction
\begin{equation}\label{eq:topPartnet}
\mathcal{L}_{\text{int}} = - \frac{m_{T}}{v}h\overline{T}T.
\end{equation}

Assuming that the top partner is a color triplet with $Q = 2/3$, then the Higgs couplings to $gg$ will receive an additional contribution from the fermion loop in Figure \ref{fig1} with $T$ running in the loop. This additional contribution is simply found by replacing the mass of the top with $m_{T}$ in the fermion loop function. Hence the deviations are readily found
\begin{equation}
\delta_{g1} = \frac{c_{g1}^{\text{SM}}(m_{t} \rightarrow m_{T})}{c_{g1}^{\text{SM}}}.
\end{equation}

 The fermion loop functions saturate quickly for large arguments, and since $m_{T}$ is constrained to be $\gg m_{t}$, the unitarity-violating scale quickly saturates, yielding the following bound which is independent of $m_{T}$
\begin{eqnarray}\label{eq:topPartner_Emax}
E^{\text{max}}_{gg} \gtrsim 11 \hspace{1mm} \text{TeV}.
\end{eqnarray}
%
\section{Conclusions}
\label{sec:conclusions}
In this paper, we investigated the scale of unitarity violation that arises from the modification of the Higgs couplings to $gg$. The unitarity of the SM at high energy relies on delicate cancelations among the various higher-order operators. This implies that any deviation in these couplings from the SM predictions would upset these cancelations, which in turn would lead to processes that have amplitudes that grow with energy, which would eventually violate unitarity at some high energy scale, signaling the onset of new physics.

In this paper, we focused on the couplings $h^{n}gg$ with $n\geq 1$. In the SM, $hgg$ is known theoretically to percent level, whereas experimentally it is only constrained to be $\sim O(0.1)$. In addition, the other couplings with $n > 1$ have neither been calculated theoretically nor been measured experimentally. This leaves plenty of room for new physics BSM. Experiments in the future, including the HL-LHC and the future $100$ TeV collider, can help probe these couplings.

We found the unitarity-violating scale in the $h^{n}gg$ to be generally higher than what was found in \cite{Chang:2019vez, Abu-Ajamieh:2020yqi} from the Higgs couplings to the top quark and $W/Z$, in addition to the Higgs trilinear coupling. This is to be expected given the fact that these couplings are loop-induced and thus are much weaker than the former tree-level ones. On the other hand, we found that the bounds are comparable or even stronger than the bounds found in \cite{Abu-Ajamieh:2021egq} from the couplings $h\gamma\gamma$ and $h\gamma Z$, which is also expected given that all these couplings are loop induced, and that $\alpha_{s} \gg \alpha$. Specifically, we found that the current level of constraints on $hgg$ allows for new physics as low as $\sim 26$ TeV. On the other hand, since the couplings $h^{n}gg$ for $ n \geq 2$ are essentially unconstrained, the scale on new physics could much lower. For instance, with the conservative assumption that $|\delta_{g n}| \leq 1$, the scale of new physics could be as low as $\sim 10$ TeV.

We also found that from the unitarity argument alone, we can both make quantitative statements about the accuracy of SMEFT, and indirectly set limits on the various couplings, especially the ones that are difficult to measure in colliders. For example, the HL-LHC is projected to measure the coupling $hgg$ at a $\pm 2.5\%$ level at $1$-$\sigma$; if this coupling is found to conform to the SM predictions to that level, then the scale of new physics from this operator is pushed to around $69$ TeV, which in turn will place stringent constraints on the coupling $hhgg$, as it will be constrained to be within $\pm 1\%$ of the SMEFT predictions.

Our main conclusion is that the current level of measurement of the Higgs properties leaves ample room for new physics BSM, whether through direct or indirect searches. A completely model-independent bottom-up approach can help us probe the scale of new physics from unitarity considerations alone. We found that while the scale of new physics in the $hgg$ is mostly beyond the reach of the LHC, it is well within the reach of the future $100$ TeV collider. In addition, accurate measurements of the Higgs properties at lower energies, especially at the HL-LHC, can help both determine the scale of new physics and place stringent limits on its couplings. 

\section*{Acknowledgments}
I would like to thank Spencer Chang for answering my questions. This work was supported by the C.V. Raman fellowship from CHEP at IISc.

\appendix
\setcounter{section}{0}

\newpage
\section{Results}
\label{app:A5}
Here we present the leading high-energy behavior for the processes used in the main text. All massive gauge bosons are understood to be longitudinally polarized unless expressly indicated otherwise. We use $-(+)$ to denote LH (RH) gluons. All processes of the form $g_{\pm}g_{\mp}X \rightarrow Y$ or $g_{\pm} X \rightarrow g_{\pm} Y$ have vanishing amplitudes as they do not conserve angular momentum. All other processes not listed in the tables are either related to the ones listed in the tables via charge conjugation or are vanishing. All amplitudes are calculated in the contact approximation, and all particles are assumed to be massless. Processes containing four gluons were calculated through summing and averaging over initial and final spins to simplify the calculation.

\begin{table}[!ht]
\centerline{
\begin{minipage}{0.8\textwidth}
\centering
\vspace{1 mm}
\tabcolsep3pt\begin{tabular}{|c|c||c|c|}
\hline
\textbf{Process} & \textbf{$\times \alpha_{s}\frac{c^{\text{SM}}_{g 1} |\delta_{g 1}|}{96\pi^{2}v^{2}}E^{2}$} &  \textbf{Process} & \textbf{$\times \alpha_{s}\frac{c^{\text{SM}}_{g 1} |\delta_{g 1}|}{96\pi^{2}v^{2}}E^{2}$} \\
\hline
$g_{\pm}g_{\pm} \rightarrow W^{+}W^{-}$ & $\sqrt{2N_{c}}$ & $g_{\pm} W^{+} \rightarrow g_{\mp} W^{+}$ & $1$\\ 
$g_{\pm}g_{\pm} \rightarrow ZZ$ & $\sqrt{N_{c}}$ & $g_{\pm} Z\rightarrow g_{\mp} Z$ & $1$\\ 
\hline
\textbf{Process} & \textbf{$\times \alpha_{s}\frac{c^{\text{SM}}_{g 1} |\delta_{g 2}|}{96\pi^{2}v^{2}}E^{2}$}  &  \textbf{Process} & \textbf{$\times \alpha_{s}\frac{c^{\text{SM}}_{g 1} |\delta_{g 2}|}{96\pi^{2}v^{2}}E^{2}$} \\
\hline
$g_{\pm}g_{\pm} \rightarrow hh$ & $\sqrt{N_{c}}$ & $g_{\pm} h\rightarrow g_{\mp} h$ & $1$\\ 
\hline
\end{tabular}
\caption{\label{tab:1} \small $|\hat{\mathscr{M}}|$ of the 4-body model-independent unitarity-violating processes arising from the modification of the Higgs coupling to $gg$. }
\end{minipage}}
\end{table}

\begin{table}[!ht]
\centerline{
\begin{minipage}{0.8\textwidth}
\centering
\vspace{1 mm}
\tabcolsep3pt\begin{tabular}{|c|c||c|c|}
\hline
\textbf{Process} & \textbf{$\times \alpha_{s} \frac{c^{\text{SM}}_{g 1} |\delta_{g 2} - \delta_{g 1}|}{1152 \sqrt{2}\pi^{3}v^{3}} E^{3}$} &  \textbf{Process} & \textbf{$\times \alpha_{s} \frac{c^{\text{SM}}_{g 1} |\delta_{g 2} - \delta_{g 1}|}{1152 \sqrt{2}\pi^{3}v^{3}} E^{3}$} \\
\hline
$g_{\pm}g_{\pm} \rightarrow hZ^{2}$ & $3\sqrt{N_{c}}$ & $g_{\pm}g_{\pm} h\rightarrow Z^{2} $ & $\sqrt{N_{c}}$\\ 
$g_{\pm}g_{\pm} Z \rightarrow hZ$ & $\sqrt{2N_{c}}$ & $g_{\pm} Z\rightarrow g_{\mp} h Z $ & $2$\\ 
$g_{\pm}h \rightarrow g_{\mp} Z^{2}$ & $\sqrt{2}$ & $g_{\pm}g_{\pm} \rightarrow hW^{+}W^{-}  $ & $3\sqrt{2N_{c}}$\\ 
$g_{\pm}g_{\pm}h \rightarrow W^{+}W^{-}$ & $\sqrt{2N_{c}}$ & $g_{\pm}g_{\pm} W^{+}\rightarrow h W^{+}  $ & $\sqrt{2N_{c}}$\\ 
$g_{\pm}h \rightarrow g_{\mp} W^{+}W^{-}$ & $2$ & $g_{\pm}W^{+}\rightarrow g_{\mp}  h W^{+}  $ & $2$\\ 
\hline
\textbf{Process} & \textbf{$\times \alpha_{s} \frac{ c^{\text{SM}}_{g1} |\delta_{g 3}| }{1152 \sqrt{2}\pi^{3}v^{3}} E^{3}$} & \textbf{Process} & \textbf{$\times \alpha_{s} \frac{ c^{\text{SM}}_{g1} |\delta_{g 3}| }{1152 \sqrt{2}\pi^{3}v^{3}} E^{3}$} \\
\hline
$g_{\pm}g_{\pm} \rightarrow h^{3}$ & $\sqrt{3N_{c}}$ &
$g_{\pm}g_{\pm} h \rightarrow h^{2}$ & $\sqrt{N_{c}}$\\ 
$g_{\pm} h \rightarrow g_{\mp} h^{2}$ & $\sqrt{2}$ & &\\ 
\hline
\end{tabular}
\caption{\label{tab:2} \small $|\hat{\mathscr{M}}|$ of the 5-body unitarity-violating processes arising from the modification of the Higgs coupling to $gg$.
}
\end{minipage}}
\end{table}

\begin{table}[!ht]
\centerline{
\begin{minipage}{0.8\textwidth}
\centering
\vspace{1 mm}
\tabcolsep3pt\begin{tabular}{|c|c|}
\hline
\textbf{Process} & \textbf{$\times \alpha_{s} \frac{ c^{\text{SM}}_{g1} |\delta_{g 2} - \delta_{g 1}|}{9216 \pi^{4}v^{4}} E^{4}$} \\
\hline
$g_{\pm} Z^{2}\rightarrow g_{\mp} Z^{2}$ & $2$ \\
$g_{\pm}g_{\pm} Z\rightarrow Z^{3} $ & $\sqrt{3N_{c}}$\\ 
$g_{\pm} ZW^{+}\rightarrow g_{\mp} ZW^{+}$ & $\frac{4}{3}$ \\
$g_{\pm} W^{+}W^{-} \rightarrow g_{\mp}Z^{2} $ & $\frac{2\sqrt{2}}{3}$\\
$g_{\pm} g_{\pm} Z \rightarrow ZW^{+}W^{-} $ & $\sqrt{2N_{c}}$\\
$g_{\pm} g_{\pm}W^{+}\rightarrow Z^{2}W^{+}$ & $\sqrt{N_{c}}$ \\
$g_{\pm} W^{+}W^{-} \rightarrow g_{\mp} W^{+}W^{-} $ & $\frac{8}{3}$\\
$g_{\pm} W^{+}W^{+} \rightarrow g_{\mp} W^{+}W^{+} $ & $\frac{4}{3}$ \\
 $g_{\pm} g_{\pm}W^{+}\rightarrow W^{+}W^{-}W^{+}$ & $2\sqrt{N_{c}}$ \\
\hline
\textbf{Process} & \textbf{$\times \alpha_{s} \frac{ c^{\text{SM}}_{g1}|\delta_{g1} - \delta_{g2}+\frac{1}{2}\delta_{g3}|}{6912 \pi^{4}v^{4}} E^{4}$} \\
\hline
$g_{\pm} hW^{+}\rightarrow g_{\mp} hW^{+}$ & $2$ \\
$g_{\pm} h^{2}\rightarrow g_{\mp} W^{+}W^{-}$ & $\sqrt{2}$ \\
$g_{\pm} g_{\pm} h\rightarrow h W^{+}W^{-}$ & $\frac{3\sqrt{N_{c}}}{\sqrt{2}}$ \\
$g_{\pm} g_{\pm} W^{+}\rightarrow h^{2} W^{+}$ & $\frac{3\sqrt{N_{c}}}{2}$ \\
$g_{\pm} hZ \rightarrow g_{\mp} h Z$ & $2$ \\
$g_{\pm} h^{2} \rightarrow g_{\mp} Z^{2}$ & $1$ \\
$g_{\pm} g_{\pm} h\rightarrow h Z^{2}$ & $\frac{3\sqrt{N_{c}}}{2}$ \\
$g_{\pm} g_{\pm} Z\rightarrow h^{2} Z$ & $\frac{3\sqrt{N_{c}}}{2}$ \\
\hline
\textbf{Process} & \textbf{$\times \alpha_{s} \frac{ c^{\text{SM}}_{g 1} |\delta_{g 4}|}{13824 \pi^{4}v^{4}} E^{4}$} \\
\hline
$g_{\pm} h^{2}\rightarrow g_{\mp} h^{2}$ & $1$ \\
$g_{\pm} g_{\pm} h\rightarrow h^{3}$ & $\frac{\sqrt{3N_{c}}}{2}$ \\
\hline
\end{tabular}
\caption{\label{tab:3} \small $|\hat{\mathscr{M}}|$ of the 6-body unitarity-violating processes arising from the modification of the Higgs coupling to $gg$.
}
\end{minipage}}
\end{table}

\begin{table}[!ht]
\centerline{
\begin{minipage}{0.8\textwidth}
\centering
\vspace{1 mm}
\tabcolsep3pt\begin{tabular}{|c|c|}
\hline
\textbf{Process} & \textbf{$\times \alpha_{s} c^{\text{SM}}_{g1} |\delta_{g 2n}| \Big( \frac{E}{4\pi v}\Big)^{2n}$} \\
\hline
$g_{\pm} g_{\pm}  h^{n-1}\rightarrow h^{n+1}$ & $\frac{\sqrt{2N_{c}}}{3(n+1)!n!\sqrt{(n+1)!(n-1)!}}$ \\
$g_{\pm} h^{n}\rightarrow  g_{\mp} h^{n}$ & $\frac{2}{3(n+1)!(n+1)!(n-1)!}$ \\
\hline
\end{tabular}
\caption{\label{tab:4} \small $|\hat{\mathscr{M}}|$ of the $2n+2$-body model-independent unitarity-violating processes arising from the modification of the Higgs coupling to $gg$.}
\end{minipage}}
\end{table}

\begin{table}[!ht]
\centerline{
\begin{minipage}{0.8\textwidth}
\centering
\vspace{1 mm}
\tabcolsep3pt\begin{tabular}{|c|c||c|c|}
\hline
\textbf{Process} & \textbf{$\times \alpha_{s}^{2} \frac{ c_{g1}^{\text{SM}}|\delta_{g1}|}{8 \pi^{2}v} (f^{abc})^{2} E$} & \textbf{Process} & \textbf{$\times \alpha_{s}^{2} \frac{ c_{g1}^{\text{SM}}|\delta_{g1}|}{8 \pi^{2}v} (f^{abc})^{2} E$} \\
\hline
$gg\rightarrow ggh$ & $\frac{N_{c}}{\sqrt{2}}$ & $gh \rightarrow g^{3}$ & $\frac{\sqrt{N_{c}}}{\sqrt{3}}$\\
\hline
\textbf{Process} & \textbf{$\times \alpha_{s}^{2} \frac{ c_{g1}^{\text{SM}}|\delta_{g1}|}{64 \pi^{3}v^{2}} (f^{abc})^{2} E^{2}$} & \textbf{Process} & \textbf{$\times \alpha_{s}^{2} \frac{ c_{g1}^{\text{SM}}|\delta_{g1}|}{64 \pi^{3}v^{2}} (f^{abc})^{2} E^{2}$} \\
\hline
$g^{2} Z \rightarrow g^{2} Z$ & $N_{c}$ & $gZ^{2} \rightarrow g^{3}$ & $\frac{\sqrt{N_{c}}}{\sqrt{3}}$ \\
$g^{2} W^{+} \rightarrow g^{2} W^{+}$ & $N_{c}$ & $gW^{-}W^{+} \rightarrow g^{3}$ & $\frac{\sqrt{2N_{c}}}{\sqrt{3}}$ \\
\hline
\textbf{Process} & \textbf{$\times \alpha_{s}^{2} \frac{ c_{g1}^{\text{SM}}|\delta_{g2}|}{64 \pi^{3}v^{2}} (f^{abc})^{2} E^{2}$} & \textbf{Process} &
 \textbf{$\times \alpha_{s}^{2} \frac{ c_{g1}^{\text{SM}}|\delta_{g2}|}{64 \pi^{3}v^{2}} (f^{abc})^{2} E^{2}$} \\
 \hline
$g^{2}h \rightarrow g^{2} h$ & $N_{c}$ & $gh^{2} \rightarrow g^{3}$ & $\frac{\sqrt{N_{c}}}{\sqrt{3}}$ \\
\hline
\end{tabular}
\caption{\label{tab:5} \small $|\hat{\mathscr{M}}|$ of the $5$- and $6$-body model-independent unitarity-violating processes from the non-derivative operators arising modification of the Higgs coupling to $gg$.}
\end{minipage}}
\end{table}

\begin{table}[!ht]
\centerline{
\begin{minipage}{0.8\textwidth}
\centering
\vspace{1 mm}
\tabcolsep3pt\begin{tabular}{|c|c|}
\hline
\textbf{Process} & \textbf{$\times \frac{\alpha_{s}^{2}}{2\pi}c_{1g}^{\text{SM}}\delta_{g(2n-2)}(f^{abc})^{2} \Big(\frac{E}{4\pi v}\Big)^{2n-2}$} \\
\hline
$g^{2}  h^{n-1}\rightarrow g^{2}  h^{n-1}$ & $\frac{N_{c}}{n!(n-1)!(n-1)!}$ \\
$g  h^{n}\rightarrow g^{3}  h^{n-2}$ & $\frac{\sqrt{2N_{c}}}{n!(n-1)!\sqrt{3 n!(n-2)!}}$ \\
$g^{4}  h^{n-3}\rightarrow h^{n+1}$ & $\frac{N_{c}}{n!(n-1)!\sqrt{6 (n+1)!(n-3)!}}$ \\
\hline
\end{tabular}
\caption{\label{tab:6} \small $|\hat{\mathscr{M}}|$ of the $2n+2$-body model-independent unitarity-violating processes from the non-derivative operators arising modification of the Higgs coupling to $gg$. Here, amplitudes were calculated by summing and averaging over initial and final spins to simplify calculations.}
\end{minipage}}
\end{table}

\end{document}